\documentclass[11pt]{cernrep}
\usepackage{graphicx}
\usepackage{bm}
\usepackage{cite,./mcite}
\usepackage{wrapfig,rotating}
\usepackage{amssymb,amsmath,array}

\begin{document}
\title{Rapidity gap survival in central exclusive diffraction: \\
Dynamical mechanisms and uncertainties\footnote{\hspace{.3em} Notice: 
Authored by Jefferson Science Associates, LLC under U.S.\ DOE
Contract No.~DE-AC05-06OR23177. The U.S.\ Government retains a
non-exclusive, paid-up, irrevocable, world-wide license to publish or
reproduce this manuscript for U.S.\ Government purposes.}}
\author{Mark Strikman$^{a}$, Christian Weiss$^{b}$}
\institute{$^{a}$ Department of Physics, Pennsylvania State University,
University Park, PA 16802, USA \\
$^{b}$ Theory Center, Jefferson Lab, Newport News, VA 23606, USA}
\maketitle
\begin{abstract}
We summarize our understanding of the dynamical mechanisms governing
rapidity gap survival in central exclusive diffraction, 
$pp \rightarrow p + H + p \; (H = \textrm{high--mass system})$, 
and discuss the uncertainties in present estimates of the
survival probability. The main suppression of diffractive
scattering is due to inelastic soft spectator interactions at small 
$pp$ impact parameters and can be described in a mean--field approximation
(independent hard and soft interactions). Moderate extra suppression 
results from fluctuations of the partonic configurations of the colliding 
protons. At LHC energies absorptive interactions of hard spectator partons
associated with the $gg \rightarrow H$ process reach the black--disk 
regime and cause substantial additional suppression, pushing the survival
probability below $0.01$.
\end{abstract}

\section{Strong interaction dynamics in rapidity gap survival}
\label{sec:1}
Calculation of the cross section of central exclusive diffraction, 
$pp \rightarrow p + H + p$ ($H$ = dijet, heavy quarkonium, Higgs boson,
\textit{etc}.) presents a major challenge for strong interaction physics.
It involves treating the hard dynamics in the elementary $gg \rightarrow H$ 
subprocess, and calculating the probability that no other interactions 
leading to hadron production occur during the $pp$ collision. 
The latter determines the suppression of diffractive relative to
non-diffractive events with the same hard process, referred to as the 
rapidity gap survival (RGS) probability. In this article we summarize
our understanding of the dynamical mechanisms determining the RGS 
probability, their phenomenological description, and the uncertainties
in present numerical predictions.

RGS in central exclusive diffraction has extensively been discussed
in an approach where soft interactions are modeled by eikonalized 
pomeron exchange; see Ref.~\cite{Khoze:2008cx} for 
a summary. More recently a partonic description was proposed, which allows 
for a model--independent formulation of the interplay of hard and soft 
interactions and reveals the essential role of the ``transverse geometry'' 
of the $pp$ collision \cite{Frankfurt:2006jp}. 
In the mean--field approximation, where hard and soft interactions are
considered as independent aside from their common dependence on the 
impact parameter, we derived a simple ``factorized'' expression for the RGS 
probability, using closure of the partonic states to take into account 
inelastic diffractive intermediate states. The resulting RGS probability
is smaller than in the models of Refs.~\cite{Khoze:2008cx,Gotsman:2008tr}
without inelastic diffraction, but comparable to the some of the versions 
of those models with multichannel diffraction. Our partonic description 
also permits us to go beyond the mean--field approximation and 
incorporate various types of correlations between the hard scattering 
process and spectator interactions. Here we discuss two such effects:
(a) quantum fluctuations of the partonic configurations of the 
colliding protons, which somewhat reduce the survival probabilities 
at RHIC and Tevatron energies; (b) absorptive interactions of high-virtuality 
spectator partons ($k^2 \sim \text{few GeV}^2$) associated with the 
hard scattering process, related to the onset of the black--disk 
regime (BDR) in hard interactions at LHC energies; this new effect 
substantially reduces the RGS probability compared to previously
published estimates.
\section{Soft spectator interactions in the mean--field approximation}
\label{sec:mean}
\begin{figure}
\begin{tabular}{lcl}
\parbox[c]{0.32\columnwidth}{
\includegraphics[width=0.32\columnwidth]{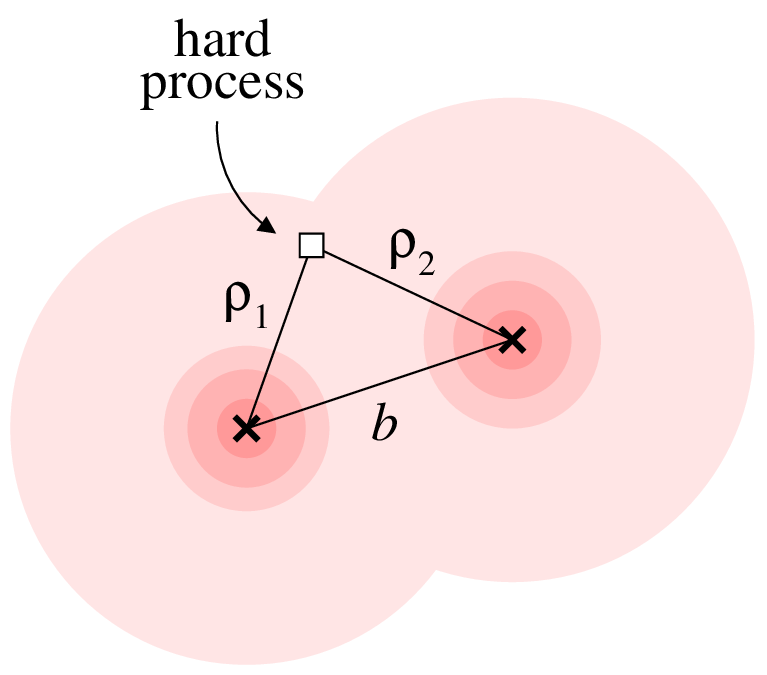}} 
& \hspace{0.1\columnwidth} &
\parbox[c]{0.49\columnwidth}{
\includegraphics[width=0.49\columnwidth]{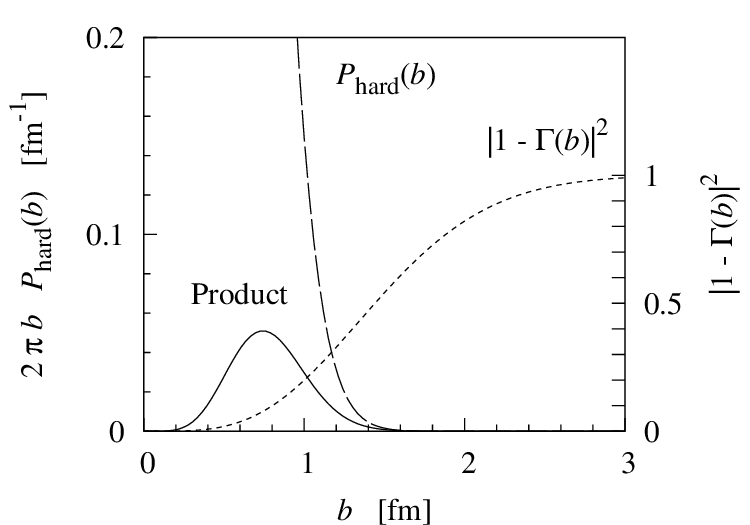}}
\\[-3ex]
{\small (a)} & & {\small (b)}
\end{tabular}
\caption[]{(a) Transverse geometry of hard diffractive $pp$ scattering.
(b) RGS probability in the impact parameter representation
\textit{cf.}\ Eq.~(\ref{survb}), for
$\sqrt{s} = 14\, \textrm{TeV}, M_H \sim 100\, \textrm{GeV}$ 
\cite{Frankfurt:2006jp}. 
Dashed line: Probability for hard scattering process 
$P_{\textrm{hard}} (b)$ (left vertical axis). Dotted line: Probability 
for no inelastic interactions between the protons, $|1 - \Gamma (b)|^2$
(right vertical axis). Solid line: Product $P_{\textrm{hard}} (b) 
|1 - \Gamma (b)|^2$ (left vertical axis).
The RGS probability Eq.~(\ref{survb}) is given by the area under this 
curve.}
\label{Fig:rgs}
\end{figure}

A simple picture of RGS is obtained 
in the impact parameter representation. On one hand, to produce the 
heavy system $H$ two hard gluons from each of the two protons need 
to collide in the same space--time point (actually, an
area of transverse size $\sim 1/ \langle k_T^2 \rangle$ in the hard 
process); because such gluons are concentrated around the 
transverse centers of the protons this is most likely when 
the protons collide at a small impact parameters, 
$b \lesssim 1 \, \text{fm}$. On the other hand, soft inelastic 
spectator interactions are strongest at small $b$ and would favor
collisions at $b \gg 1 \, \text{fm}$ for diffractive scattering. 
These different preferences limit diffraction to an intermediate 
range of impact parameters and ensure that its 
cross section is substantially suppressed compared to 
non--diffractive scattering. More precisely, 
the RGS probability is given by \cite{Frankfurt:2006jp}
\begin{equation}
S^2 \;\; = \;\; 
\int d^2 b \; P_{\textrm{hard}} (b) \; |1 - \Gamma (b)|^2 ,
\hspace{4em} b \equiv |\bm{b}|.
\label{survb}
\end{equation}
Here $P_{\textrm{hard}} (b)$ is the probability for two 
gluons to collide at the same transverse point as a function of
the $pp$ impact parameter, given by the convolution of the 
transverse spatial distributions of the gluons in the colliding protons, 
normalized to $\int d^2 b \; P_{\textrm{hard}} (b) = 1$ 
(see Fig.~\ref{Fig:rgs}a).
The factor $|1- \Gamma(b)|^2$ is the probability for the two protons 
not to interact inelastically in a collision at the given impact parameter,
calculable in terms of the profile function of the $pp$ elastic 
amplitude, $\Gamma(b)$. Figure~\ref{Fig:rgs}b shows the $b$--dependence
of the two factors as well as their product, illustrating the
interplay described above. 
While we have motivated Eq.~(\ref{survb}) by probabilistic arguments,
it actually can be derived (as well as the expression for the
differential cross section) in the partonic description 
of Ref.~\cite{Frankfurt:2006jp} within the mean--field approximation, 
where one assumes no correlation between the presence of the gluons 
involved in the hard interaction (with the particular $x$) and the 
strength of the soft spectator interactions. In this approximation 
one can use closure to sum over the different diffractive intermediate states, 
and thus effectively include the contribution of inelastic 
diffraction.\footnote{In principle there is also a contribution
from excitation of a diffractive state by soft spectator interactions 
and subsequent transition back to the proton via the nondiagonal
gluon GPD; however, it is strongly suppressed because the typical 
excitation masses in hard and soft diffraction are very different
in the kinematics of Higgs production at the LHC
($10^{-8}\le x_{I\!\!P}\le 0.1$ for generic $pp$ diffraction and
$10^{-2}\le x_{I\!\!P}\le 0.1$ for the GPD); see Section IV C of
Ref.~\cite{Frankfurt:2006jp}.} The numerical values of
the RGS probability obtained from Eq.~(\ref{survb}) are of the 
order $S^2 \sim 0.03$ for $M_H = 100\, \text{GeV}$ and 
$\sqrt{s} = 14 \, \text{TeV}$; see Ref.~\cite{Frankfurt:2006jp} 
for details.
%
%
%

%
%
\begin{figure}
\parbox[c]{.5\textwidth}{
\includegraphics[width=.5\textwidth]{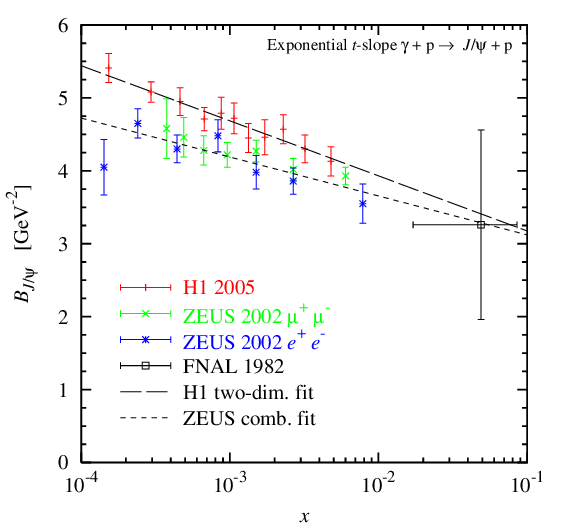}}
\hspace{.05\textwidth}
\parbox[c]{.4\textwidth}{
\caption[]{The exponential $t$--slope, $B_{J/\psi}$, 
of exclusive $J/\psi$ photoproduction, extracted from fits to
the FNAL E401/E458 \cite{Binkley:1981kv}, HERA H1 \cite{Aktas:2005xu}, 
and ZEUS \cite{Chekanov:2004mw} data. The long--dashed and 
short--dashed lines represent fits to the $x$--dependence of
the H1 and ZEUS $t$--slopes \cite{Aktas:2005xu,Chekanov:2004mw}, 
\textit{cf.}\ Eq.~(\ref{B_x}). The $t$--slope of the gluon GPD,
$B_g$, is obtained from $B_{J/\psi}$ after applying a small
correction for the finite size of the $J/\psi$ \cite{Frankfurt:2005mc}.} 
\label{fig:bpsi}}
\end{figure}
It is worthwhile to discuss the uncertainty in the numerical predictions 
for $S^2$ in the mean--field approximation, Eq.~(\ref{survb}), resulting
from our imperfect knowledge of the functions in the integrand. 
We first consider the transverse spatial distribution of gluons entering 
in $P_{\textrm{hard}}(b)$. The latter is obtained as the Fourier 
transform of the $t$--dependence (more precisely, transverse momentum
dependence) of the gluon generalized parton distribution (GPD) measured 
in hard exclusive vector meson production. Extensive studies at HERA
have shown that exclusive $J/\psi$ photoproduction, $\gamma p \rightarrow
J/\psi + p$, provides an effective means for probing the 
$t$--dependence of the gluon GPD at small and intermediate $x$
(a small correction for the finite transverse size of the $J/\psi$ 
is applied) \cite{Frankfurt:2005mc}. Figure~\ref{fig:bpsi} 
summarizes the results for the exponential $t$--slope of this 
process, $B_{J/\psi}$, from HERA H1 \cite{Aktas:2005xu} and 
ZEUS \cite{Chekanov:2004mw} and the FNAL E401/E458 experiment
\cite{Binkley:1981kv}, as well as fits to the $x$--dependence of
the H1 and ZEUS results of the form (here $x = M_{c\bar c}^2/W^2$) 
\begin{equation}
B_{J/\psi}(x) \;\; = \;\; B_{J/\psi}(x_0) \; + \; 2 \alpha'_{J/\psi}
\; \ln (x_0/x) .
\label{B_x}
\end{equation}
There is a systematic difference between the H1 and ZEUS results due 
to different analysis methods \cite{Aktas:2005xu,Chekanov:2004mw}; 
however, the fits to both sets agree well with the FNAL
point when extrapolated to larger $x$. In diffractive production
of a system with $M_H = 100\, \text{GeV}$ at $\sqrt{s} = 14\, \textrm{TeV}$ 
at zero rapidity the gluons coupling to the heavy system $H$ have
momentum fractions $x_{1, 2} = M_H/ \sqrt{s} = 0.007$. Assuming 
exponential $t$--dependence of the gluon GPD, we can estimate the 
uncertainty in the transverse spatial distribution of gluons 
at such $x$ by evaluating the fits to the HERA
data within the error bands quoted for $B_{J/\psi}(x_0)$ and 
$\alpha'_{J/\psi}$ \cite{Aktas:2005xu,Chekanov:2004mw}. We find a 
15-20\% uncertainty of $B_{J/\psi}$ at $x = 0.007$ in this way,
translating into a 20--30\% uncertainty in the mean--field 
RGS probability, 
Eq.~(\ref{survb}). We note that there is at least a comparable 
uncertainty in $S^2$ from the uncertainty of the shape of the 
$t$--dependence; this is seen from Fig.~10 of 
Ref.~\cite{Frankfurt:2006jp}, where the 
exponential is compared with a theoretically motivated dipole form 
which also describes the FNAL data. 
Altogether, we estimate that our imperfect knowledge of the spatial 
distribution of gluons results in an uncertainty of the mean--field 
result for $S^2$ by a factor $\sim 2$. Dedicated 
analysis of the remaining HERA exclusive data, and particularly precision 
measurements with a future electron--ion collider (EIC), could 
substantially improve our knowledge of the transverse spatial 
distribution of gluons.

We now turn to the uncertainty in $S^2$ arising from the $pp$ elastic 
amplitude, $\Gamma (b)$. Most phenomenological analyses of $pp$ 
elastic and total cross section data find that for TeV energies
$|1 - \Gamma (b)| \le  0.05$ at $b = 0$, corresponding to 
near--unit probability of inelastic interactions at small impact 
parameters (BDR). This is supported
by theoretical studies in the QCD dipole model,
which show that the large--$x$ partons with virtualities of up to several 
$\text{GeV}^2$ experience ``black'' interactions with the small--$x$ 
gluon field in the other proton when passing through the other proton
at transverse distances $\rho\le 0.5 \text{fm}$, and 
receive transverse momenta $k_T \ge 1 \, \textrm{GeV}$
(see Ref.~\cite{Frankfurt:2005mc} for a summary). 
At $pp$ impact parameter $b = 0$
the chance that none of the leading partons in the protons receive such 
a kick is extremely small, implying that 
$|1 - \Gamma (b)| \sim 0$ \cite{Frankfurt:2004fm}.
For the RGS probability in the mean--field approximation, Eq.~(\ref{survb}),
the fact that $|1 - \Gamma (b)|^2$ is small at $b = 0$ is essential, 
as this eliminates the contribution from small $b$ in the integral
(see Fig.~\ref{Fig:rgs}b)
and stabilizes the numerical predictions. However, present theoretical 
arguments and data analysis cannot exclude a small non-zero value of 
$|1 - \Gamma (b)|$ at $b = 0$; a recent analysis finds 
$|1 - \Gamma (b)| \sim 0.1$ \cite{Luna:2008pp}. 
To investigate the potential 
implications for the RGS probability, we evaluate Eq.~(\ref{survb})
with the Gaussian parametrization of $\Gamma (b)$ 
of Ref.~\cite{Frankfurt:2006jp},
Eq.~(12), but with $\Gamma (b = 0) = 1 - \epsilon$. 
We find that a value of $\epsilon = 0.1$, corresponding to 
$|1 - \Gamma (b)|^2 = 0.01$, increases the mean--field result
for $S^2$ by a factor $\sim 1.8$, indicating significant uncertainty
of the mean--field result. However, as explained in Sec.~\ref{sec:BDR_hard}
below, hard spectator interactions associated with the $gg \rightarrow H$
process lead to an additional suppression of diffraction at small $b$
(not contained in the soft RGS probability), which mitigates the impact
of this uncertainty on the overall diffractive cross section.
\section{Fluctuations of parton densities and soft--interaction strength}
Corrections to the mean--field picture of RGS arise from 
fluctuations of the interacting configurations in the colliding protons. 
This concept is known well in soft diffraction, where fluctuations of 
the strength of interaction between the colliding hadrons give rise 
to inelastic diffraction. In hard diffraction, one expects that 
also the gluon density fluctuates; \textit{e.g.}\ because the color
fields are screened in configurations of small 
size \cite{Frankfurt:2008vi}. In fact, the
variance of the gluon density fluctuations can be directly related 
to the ratio of inelastic and elastic diffraction in processes 
such as $\gamma^*_L + p \rightarrow \textrm{``vector meson''} + X$,
\begin{equation}
\omega_g \;\; \equiv \;\; 
\frac{\langle G^2 \rangle - \langle G \rangle^2}{\langle G \rangle^2}
\;\; = \;\; 
\left. \frac{d\sigma_{\text{inel}}}{dt} \right/ \left.
\frac{d\sigma_{\text{el}}}{dt} \right|_{t=0} .
\label{omega_g}
\end{equation}
The HERA data are consistent with the dynamical model estimate 
of $\omega_g \sim 0.15-0.2$ for $Q^2 = 3 \, \text{GeV}^2$ and 
$x \sim 10^{-4} - 10^{-3}$ \cite{Frankfurt:2008vi}; 
unfortunately, the limited $Q^2$ range 
and the lack of dedicated studies do not allow for a more precise
extraction of this fundamental quantity.

In central exclusive diffraction, correlated fluctuations of the 
soft--interaction strength and the gluon density lower 
the RGS probability, because small-size configurations which 
experience little absorption have a lower gluon density.
This effect can be modeled by a generalization of the mean--field
expression (\ref{survb}), in which both the gluon GPDs in $P_{\text{hard}}$
and the profile function fluctuate as a function of an external 
parameter controlling the overall size of the 
configurations \cite{Frankfurt:2008vi}.
Numerical studies find a reduction of the RGS probability by a factor
$\sim 0.82 \, (0.74)$ for a system with mass $M_H = 100 \, \text{GeV}$ 
produced at zero rapidity at 
$\sqrt{s} = 2 \, (14) \, \textrm{TeV}$. The dynamical model
used in this estimate does not include fluctuations of the gluon density 
at larger $x (\sim 0.05 - 0.1)$, 
which could increase the suppression.

We emphasize again that inelastic diffraction \textit{per se} 
is included in the partonic approach of Ref.~\cite{Frankfurt:2006jp} 
through the closure of partonic states. The effect discussed
in this section is specifically related to correlations between 
the fluctuations of the parton densities and the soft--interaction strength;
in the limit of zero correlations (independent fluctuations) 
we recover the mean--field result described 
above \cite{Frankfurt:2008vi}.
 
\section{Black--disk regime in hard spectator interactions}
\label{sec:BDR_hard}
Substantial changes in the mechanism of diffractive scattering are 
brought about by the onset of the BDR in hard interactions at LHC 
energies, where even highly virtual partons ($k^2 
\sim \textrm{few GeV}^2$) with $x \gtrsim 10^{-2}$ experience 
``black'' interactions with the small--$x$ gluons in the 
other proton. This new effect modifies the amplitude of
central exclusive diffraction in several ways: (a)
absorption of the ``parent'' partons of the gluons attached 
to the high--mass system; (b) absorption of the hard gluons 
attached to the high--mass system; (c) absorption due to local 
interactions within the partonic ladder. Such absorptive hard
interactions cause additional suppression of diffractive 
scattering, not included in the traditional soft--interaction 
RGS probability \cite{Frankfurt:2006jp}. Because of the 
generic nature of ``black'' interactions, we can estimate
this effect by a certain modification of the mean--field picture
in the impact parameter representation. Here we focus on mechanism (a)
and show that it causes substantial suppression; the other 
mechanisms may result in further suppression.

%
%
\begin{figure}
\includegraphics[width=0.98\textwidth]{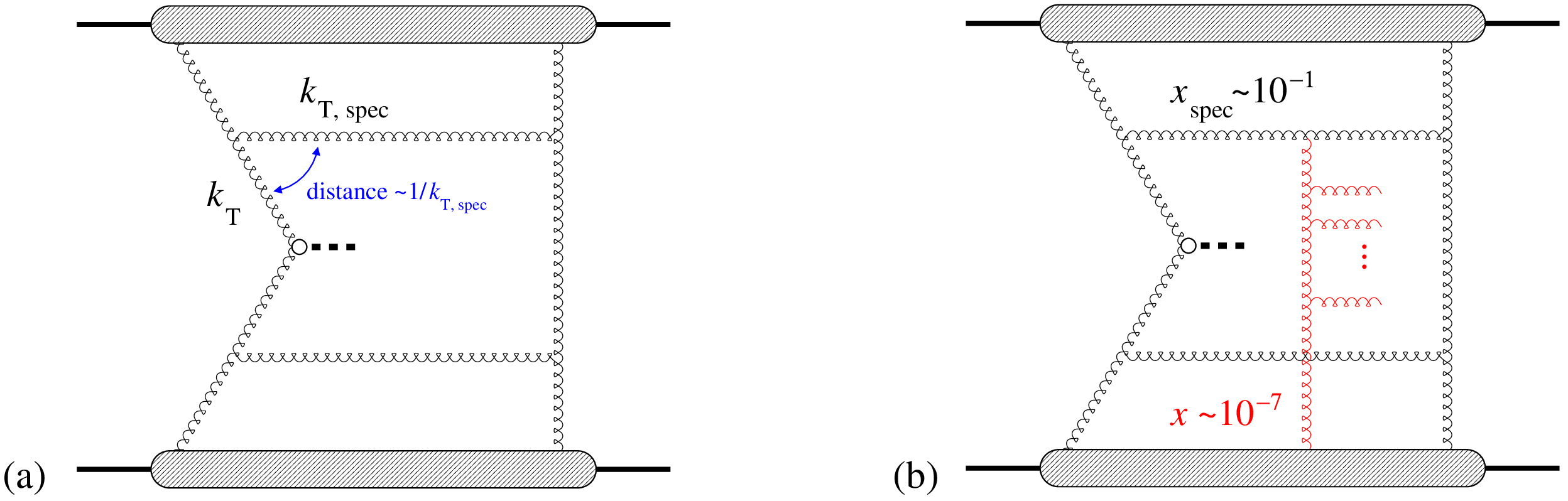}
\caption{(a) QCD evolution--induced correlation between hard partons.
The transverse distance between the active parton and the spectator
is $\sim 1/k_{T, \, {\rm spec}}$.
(b) Absorptive interaction of the hard spectator with small--$x$
gluons in the other proton.}
\label{hardabs}
\end{figure}
According to Ref.~\cite{Khoze:2000cy} (and references therein)
the dominant contribution to the hard amplitude of Higgs production 
at the LHC ($M_H = 100 \, \text{GeV}, x_{1, 2} \sim 10^{-2}$) 
originates from gluons with transverse momenta of the order 
$k_T \sim 2\,\text{GeV}$. 
Such gluons are typically generated by DGLAP evolution starting
from the initial scale, $Q^2_0$, in which spectator partons,
mostly gluons, are emitted (see Fig.\ref{hardabs}a). 
In the leading--log approximation $Q_0 \ll k_{T, \, {\rm spec}} \ll k_T$, 
and thus the transverse distance between the active and spectator 
parton is $\sim 1/k_{T, \, {\rm spec}} \ll R_{\rm proton}$, 
amounting to short--range correlations between partons. 
If the interactions of the spectator parton with the small--$x$ 
gluons in the other proton become significant (see Fig.\ref{hardabs}b), 
the basic assumption of the mean--field approximation --- that the 
spectator interactions are independent of the hard process --- 
is violated, and the interactions of that parton need to be
treated separately. Indeed, studies within the QCD dipole 
model show that at the LHC energy spectator gluons with 
$k_{T, \, {\rm spec}} \sim 1 \, \text{GeV}$ and $x_{\rm spec} \sim 10^{-1}$
``see'' gluons with momentum fractions $x \sim 10^{-7}$ in the other proton, 
and are absorbed with near--unit probability if their impact parameters 
with the other proton are less than $\sim 1 \, \textrm{fm}$
\cite{Frankfurt:2006jp}.\footnote{The cross section of ``gluonic'' ($88$) 
dipoles is larger
than that of the quark--antiquark ($\bar 3 3$) dipoles in $\gamma^\ast p$
scattering \cite{Rogers:2003vi} by a factor $9/4$. A summary plot
of the profile function for gluon--proton scattering is given 
in Fig.~13 of Ref.~\cite{Frankfurt:2005mc} (right $y$--axis). Note that 
$\Gamma^{\rm gluon-proton} = 0.5$ already corresponds to a significant
absorption probability of $1 - |1 - \Gamma^{\rm gluon-proton}|^2 = 0.75$.} 
For $pp$ impact parameters 
$b < 1 \, \textrm{fm}$ about $90\%$ of the strength in 
$P_{\textrm{hard}}(b)$ comes from parton--proton impact parameters 
$\rho_{1, 2} < 1 \, \textrm{fm}$ (\textit{cf}.\ Fig.~\ref{Fig:rgs}a),
so that this effect practically eliminates diffraction at 
$b < 1 \, \textrm{fm}$. Since $b < 1 \, \textrm{fm}$ accounts for 2/3 
of the cross section (see Fig.~\ref{Fig:rgs}b),
and the remaining contributions at $b > 1 \, \textrm{fm}$ are also 
reduced by absorption, we estimate that absorptive interactions
of hard spectators in the BDR reduce the RGS probability at LHC
to about 20\% of its mean--field value. Much less suppression
is expected at the Tevatron energy, where hard spectator
interactions only marginally reach the BDR.

In the above argument one must also allow for the possibility of 
trajectories with no gluon emission, which correspond 
to the Sudakov form factor--suppressed $\delta(1 - x)$--term in the 
evolution kernel. While such trajectories are not affected by absorption,
their contributions are small both because of the Sudakov suppression,
and because they effectively probe the gluon density at a low
scale, $Q_0^2 \sim 1 \, \textrm{GeV}^2$, where evolution--induced 
correlations between partons can be neglected. We estimate that the
contribution of such trajectories to the cross section is suppressed 
compared to those with emissions by a factor
$R = \left[ S_G^2 \, G(x, Q^2) / G(x, Q_0^2) \right]^2 \sim 1/10$, 
where $S_G^2 = \exp[-(3\alpha_s/\pi) 
\ln^2 (Q^2/Q_0^2)]$ is the square of the Sudakov form factor,
and $Q^2 \sim 4 \, \textrm{GeV}^2$. Their net contribution 
is thus comparable to that of the trajectories 
with emissions, because the latter are strongly suppressed by the 
absorption effect described above. Combining the two, we obtain
an overall suppression by a factor of the order $\sim 0.3$. 
More accurate estimates would need to take into account fluctuations 
in the number of emissions; in particular, trajectories on which 
only one of the partons did not emit gluons are suppressed
only by $\sqrt{R}$ and may make significant contributions.

The absorptive hard spectator interactions described here ``push'' 
diffractive $pp$ scattering to even larger impact parameters
than would be allowed by the soft spectator interactions included
in the mean--field RGS probability, Eq.~(\ref{survb}) (except for the 
Sudakov--suppressed contribution).
One interesting consequence of this is that it makes the uncertainty 
in the mean--field prediction arising from $\Gamma(0) \neq 1$ 
(see Sec.~\ref{sec:mean}) largely irrelevant, as the region of
small impact parameters is now practically eliminated by the 
hard spectator interactions. Another consequence is that the
final--state proton transverse momentum distribution is shifted to
to smaller values; this could in principle be observed in 
$p_T$--dependent measurements of diffraction.
We note that the estimates of hard spectator interactions reported here 
are based on the assumption that DGLAP evolution reasonably well describes 
the gluon density down to $x \sim 10^{-6}$; the details 
(but not the basic picture) may change if small--$x$ resummation 
corrections were to significantly modify the gluon density at such values 
of $x$ (see Ref.~\cite{Ciafaloni:2007gf} and references therein). 

\section{Summary}
The approach to the BDR in the interaction of hard spectator partons,
caused by the increase of the gluon density at small $x$, has profound 
implications for central exclusive diffraction at LHC:
\textit{No saturation without disintegration!} 
The RGS probability is likely to be much smaller (by a factor of $\sim 1/3$ 
or less) than predicted by the mean--field approximation or corresponding 
models which neglect correlations of partons in the transverse plane.
Diffractive scattering is relegated either to very large impact parameters
($b> 1\, \text{fm}$) or to Sudakov--suppressed trajectories without 
gluon radiation. We estimate that the overall RGS probability at LHC is
$S^2 < 0.01$. Extrapolation of the Tevatron results may be 
misleading because interactions of hard spectators are generally 
far from ``black'' at that energy. The new effects described here call 
for detailed MC--based studies of possible histories of the hard 
scattering process and their associated spectator interactions.

\bibliographystyle{heralhc} 
{\raggedright
\bibliography{heralhc}

\providecommand{\etal}{et al.\xspace}
\providecommand{\coll}{Coll.}
\catcode`\@=11
\def\@bibitem#1{%
\ifmc@bstsupport
  \mc@iftail{#1}%
    {;\newline\ignorespaces}%
    {\ifmc@first\else.\fi\orig@bibitem{#1}}
  \mc@firstfalse
\else
  \mc@iftail{#1}%
    {\ignorespaces}%
    {\orig@bibitem{#1}}%
\fi}%
\catcode`\@=12
\begin{mcbibliography}{10}

\bibitem{Khoze:2008cx}
Khoze, V. A. and Martin, A. D. and Ryskin, M. G.,
\newblock Eur. Phys. J.{} {\bf C55},~363~(2008)\relax
\relax
\bibitem{Frankfurt:2006jp}
Frankfurt, L. and Hyde, C. E. and Strikman, M. and Weiss, C.,
\newblock Phys. Rev.{} {\bf D75},~054009~(2007)\relax
\relax
\bibitem{Gotsman:2008tr}
Gotsman, E. and Levin, E. and Maor, U. and Miller, J. S.~(2008)\relax
\relax
\bibitem{Binkley:1981kv}
Binkley, M. and others,
\newblock Phys. Rev. Lett.{} {\bf 48},~73~(1982)\relax
\relax
\bibitem{Aktas:2005xu}
Aktas, A. and others,
\newblock Eur. Phys. J.{} {\bf C46},~585~(2006)\relax
\relax
\bibitem{Chekanov:2004mw}
Chekanov, S. and others,
\newblock Nucl. Phys.{} {\bf B695},~3~(2004)\relax
\relax
\bibitem{Frankfurt:2005mc}
Frankfurt, L. and Strikman, M. and Weiss, C.,
\newblock Ann. Rev. Nucl. Part. Sci.{} {\bf 55},~403~(2005)\relax
\relax
\bibitem{Frankfurt:2004fm}
Frankfurt, L. and Strikman, M. and Zhalov, M.,
\newblock Phys. Lett.{} {\bf B616},~59~(2005)\relax
\relax
\bibitem{Luna:2008pp}
Luna, E. G. S. and Khoze, V. A. and Martin, A. D. and Ryskin, M.
  G.~(2008)\relax
\relax
\bibitem{Frankfurt:2008vi}
Frankfurt, L. and Strikman, M. and Treleani, D. and Weiss, C.,
\newblock Phys. Rev. Lett.{} {\bf 101},~202003~(2008)\relax
\relax
\bibitem{Khoze:2000cy}
Khoze, Valery A. and Martin, Alan D. and Ryskin, M. G.,
\newblock Eur. Phys. J.{} {\bf C14},~525~(2000)\relax
\relax
\bibitem{Rogers:2003vi}
Rogers, T. and Guzey, V. and Strikman, M. and Zu, X.,
\newblock Phys. Rev.{} {\bf D69},~074011~(2004)\relax
\relax
\bibitem{Ciafaloni:2007gf}
Ciafaloni, M. and Colferai, D. and Salam, G. P. and Stasto, A. M.,
\newblock JHEP{} {\bf 08},~046~(2007)\relax
\relax
\end{mcbibliography}
}
\end{document}